\documentstyle[epsfig,subfig,amsmath]{elsarticle}
\newcommand{\newc}{\newcommand}
\newc{\lra}{\leftrightarrow}
\newc{\beq}{\begin{equation}}
\newc{\eeq}{\end{equation}}
\newc{\barr}{\begin{eqnarray}}
\newc{\earr}{\end{eqnarray}}

\newcommand{\beqa}{\begin{eqnarray}}
\newcommand{\eeqa}{\end{eqnarray}}
\newcommand{\bdm}{\begin{displaymath}}
\newcommand{\edm}{\end{displaymath}}

\renewcommand{\thefootnote}{\#\arabic{footnote}}
\renewcommand{\theequation}{\thesection.\arabic{equation}}
\renewcommand{\thefigure}{\thesection.\arabic{figure}}
\renewcommand{\thetable}{\thesection.\arabic{table}}

\begin{document}
\def\vbf{\mbox{\boldmath $\upsilon$}}
\def\barr{\begin{eqnarray}}
\def\earr{\end{eqnarray}}
\def\g{\gamma}
\newcommand{\dphi}{\delta \phi}
\newcommand{\bupsilon}{\mbox{\boldmath \upsilon}}
\newcommand{\bfup}{\mbox{\boldmath \upsilon}}
\newcommand{\at}{\tilde{\alpha}}
\newcommand{\pt}{\tilde{p}}
\newcommand{\Ut}{\tilde{U}}
\newcommand{\rhb}{\bar{\rho}}
\newcommand{\pb}{\bar{p}}
\newcommand{\pbb}{\bar{\rm p}}
\newcommand{\kt}{\tilde{k}}
\newcommand{\wt}{\tilde{w}}

\date{\today}
\title {Modulation, asymmetry and the diurnal variation  in axionic dark matter searches}
%
%
%
%
%
\author{Y. Semertzidis and J.D. Vergados}
%
%
\address{
 KAIST University, Daejeon, Republic of Korea and \\
Center for Axion and Precision Physics Research, IBS, Daejeon 305-701, Republic of Korea
}
\begin{frontmatter}
\begin{abstract}
In the present work we study possible time dependent effects in Axion Dark Matter searches  employing  resonant cavities. We find that the width of the resonance, which depends on the axion mean square  velocity in the local frame, will show an annual variation due to the motion of the Earth  around the sun (modulation). Furthermore, if the experiments become  directional, employing suitable resonant cavities, one expects large asymmetries in the observed widths relative to the sun's direction of motion. Due to the rotation of the Earth around its axis, these asymmetries will manifest themselves as a diurnal variation in the observed width.
\end{abstract}

\end{frontmatter}
\section{Introduction}
\begin{figure}[!ht]
\begin{center}
\includegraphics[width=\textwidth, height=\textwidth]{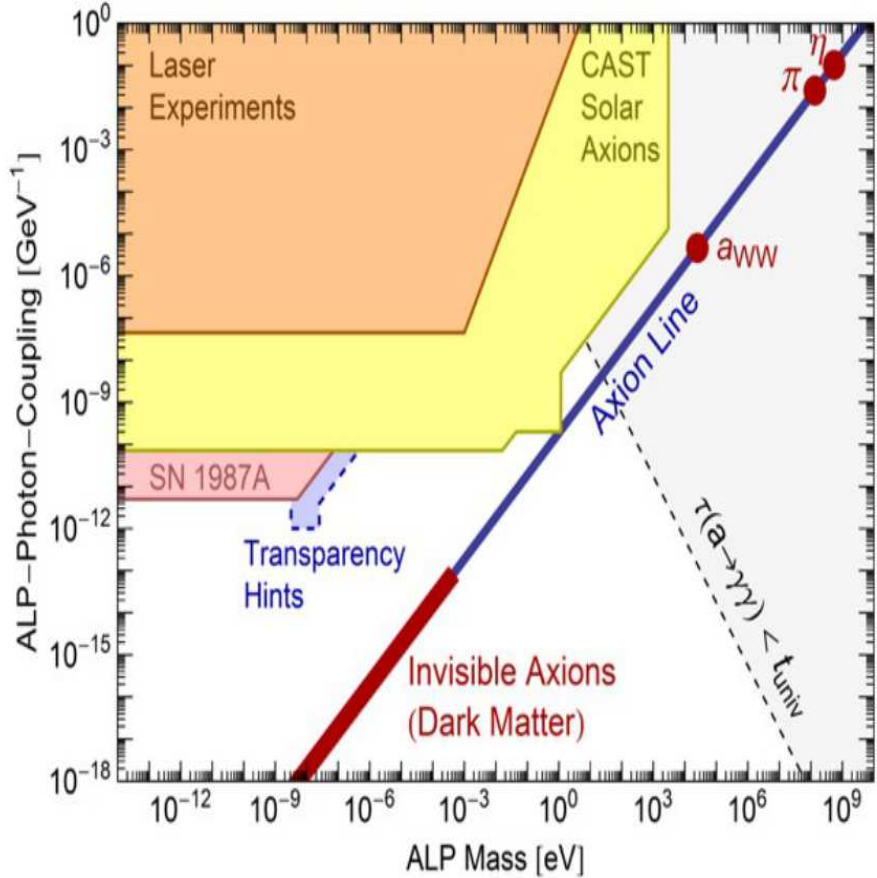}
 \caption{The  parameter space for axion like particles  (courtesy of professor Raffelt)}
 \label{fig:AxionParSp}
  \end{center}
  \end{figure}

The axion has been proposed a long time ago as a solution to the strong CP problem \cite{PecQui77} resulting to a pseudo Goldstone Boson \cite{SWeinberg78,Wilczek78,PWW83,AbSik83,DineFisc83}, but it has also been recognized as a prominent  dark matter candidate \cite{PriSecSad88}.
 In fact, realizing an idea proposed a long time ago by Sikivie \cite{Sikivie83}, various experiments such as ADMX and ADMX-HF collaborations 
\cite{Stern14,MultIBSExp}, \cite{ExpSetUp11b},\cite{ADMX10} are now planned to search for them. In addition, the newly established center for axion and physics research (CAPP)  has started an ambitious axion dark matter research program \cite{CAPP}, using SQUID and HFET technologies \cite{ExpSetUp11a}.
  The allowed parameter space has been presented in a nice slide by Raffelt \cite{MultIBSTh} in the recent Multidark-IBS workshop.  From Fig. \ref{fig:AxionParSp}  containing information for all the axion like particles, we are interested in the regime allowed for invisible axions, which can be dark matter candidates.
	
In the present work we will take the view that the axion is non relativistic with mass in $\mu$eV-meV scale moving with an average velocity which is $\approx0.8\times10^{-3}$c.  The width of the observed resonance depends on the axion mean square velocity in the local frame. Thus one expects it to exhibit a time variation due to the motion of the Earth. Furthermore in directional experiments involving long cavities, one expects asymmetries with regard to the sun's direction of motion as it goes around the center of the galaxy. Due to the rotation of the Earth around its axis these asymmetries in the width of the resonance will manifest themselves in their diurnal variation. These two special signatures, expected to be sizable,  may aid the analysis of axion dark searches in discriminating against possible backgrounds.
\section {Brief summary of the formalism}
The photon axion interaction is dictated by the Lagrangian:
\beq
{\cal L}_{a \gamma\gamma}=g_{a\gamma\gamma}a {\bf E}\cdot{\bf B},\, g_{a\gamma\gamma}=\frac{\alpha g_{\gamma}}{ \pi f_a},
\eeq
where  ${\bf E}$ and ${\bf B}$ are the electric and magnetic fields, $g_{\gamma}$ a model dependent constant of order one 
\cite{Stern14},\cite{HKNS14}\cite{JEKim98} and $f_a$ the axion decay constant. Axion dark matter detectors \cite{HKNS14} employ an external magnetic field, ${\bf B}\rightarrow {\bf B}_0$ in the previous equation, in which case one of the photons is replaced by a virtual photon, while the other maintains the energy of the axion, which is its mass plus a small fraction of kinetic energy. \\
The power produced, see e.g.  \cite{Stern14}, is given by:
\beq
P_{mnp}=g_{a\gamma\gamma}^2\frac{\rho_a}{m_a} B_0^2 V C_{mnp} Q_L
\label{Eq:Pmnp}
\eeq
$Q_L$ is the loaded quality factor of the cavity. Here we have assumed $Q_L$ is smaller than the axion width $Q_a$, see below.  More generally, $Q_L$ should be substituted by min ($Q_L$, $Q_a$).
This power depends on the axion density and is pretty much independent of the velocity distribution.
	\label{powerspecrum}
	
The axion power spectrum, which is of great interest to experiments, is written as a Breit-Wigner shape \cite{HKNS14}, \cite{KMWM85}:
\beq
\left |{\cal A}(\omega) \right |^2=\frac{ \rho_D}{m_a^2}\frac{\Gamma}{(\omega-\omega_a)^2+(\Gamma/2)^2},\Gamma=\frac{\omega_a}{Q_a}
\eeq
with $\omega_a=m_a\left (1+(1/6)\prec\upsilon^2\succ\right )$ and $Q_{a}=m_a/\left (m_a /(m_a\prec\upsilon^2\succ/3)\right ).$
The width explicitly depends on the average axion velocity squared in the laboratory. Thus the width in the laboratory is affected by  the sun's motion. In the non directional experiments $\prec\upsilon^2\succ=(3/2 )\upsilon_0^2$ becomes $\prec\upsilon^2\succ=(5/2) \upsilon_0^2$ ($\upsilon_0$ the velocity of the sun around the center of the galaxy). If we take into account the motion of the Earth around the sun the width becomes time dependent (modulation) as described below (see section  \ref{sec:mod}).\\
 The situation becomes more dramatic as soon the experiment is directional. In this case the width depends strongly on the direction of observation relative to the sun's direction of motion.
 Directional experiments can, in principle, be performed by changing the orientation of a long  cavity \cite{Stern14},\cite{FutExp14},\cite{IrasGarcia12}, provided that the axion wavelength is not larger than the length of the cylinder, $\lambda_a\le h$. In the ADMX \cite{Stern14} experiment $h=100$cm, while  from their Fig. 3 one can see that  the relevant for dark matter wavelengths $\lambda_a$ are  between 1 and 65 cm.
\section{Modification of the width due to the motion of the Earth and the sun.}
From the above discussion it appears that velocity distribution of axions may play a role in the experiments. 
\subsection{The velocity distribution}
If the axion is going to be considered as dark matter candidate, its density should fit the rotational curves.  Thus for temperatures $T$  such that $m_a/T\approx 4 \times 10^{6}$ the velocity distribution can be taken  to be   analogous to that assumed for WIMPs, i.e. a M-B distribution  with a characteristic velocity which equals the velocity of the sun around the center of the galaxy, i.e. $\upsilon_0 \approx 220$km/s. So we will employ the distribution:
\beq
f({\vec \upsilon})=\frac{1}{(\sqrt{\pi}\upsilon_0)^3}e^{-\frac{\upsilon^2}{\upsilon^2_0}}
\eeq
In order to compute the average of the velocity squared entering the power spectrum we need to find the local velocity distribution by taking into account the velocity of the Earth around the sun and the velocity of the sun around the center of the galaxy. The first motion leads to a time dependence of the observed signal in standard experiments , while the latter motion leads to asymmetries in directional experiments. 

\subsection{The annual modulation in non directional experiments}
\label{sec:mod}
The modification of the velocity distribution in the local frame due to annual motion of the Earth is expected to affect the detection of axions in a time dependent way, which, following the terminology of the standard WIMPs, will be called the modulation effect 
\cite{DFS86} (the corresponding effect due to the rotation of the Earth around its own axis is too small to be observed). Periodic signatures  for the detection of cosmic axions were first considered by Turner \cite{Turner90}.
 
\begin{figure}[!ht]
\begin{center}
\rotatebox{90}{\hspace{-0.0cm} {$y^2 f(y)\longrightarrow$}}
\includegraphics[width=0.9\textwidth,height=0.3\textwidth]{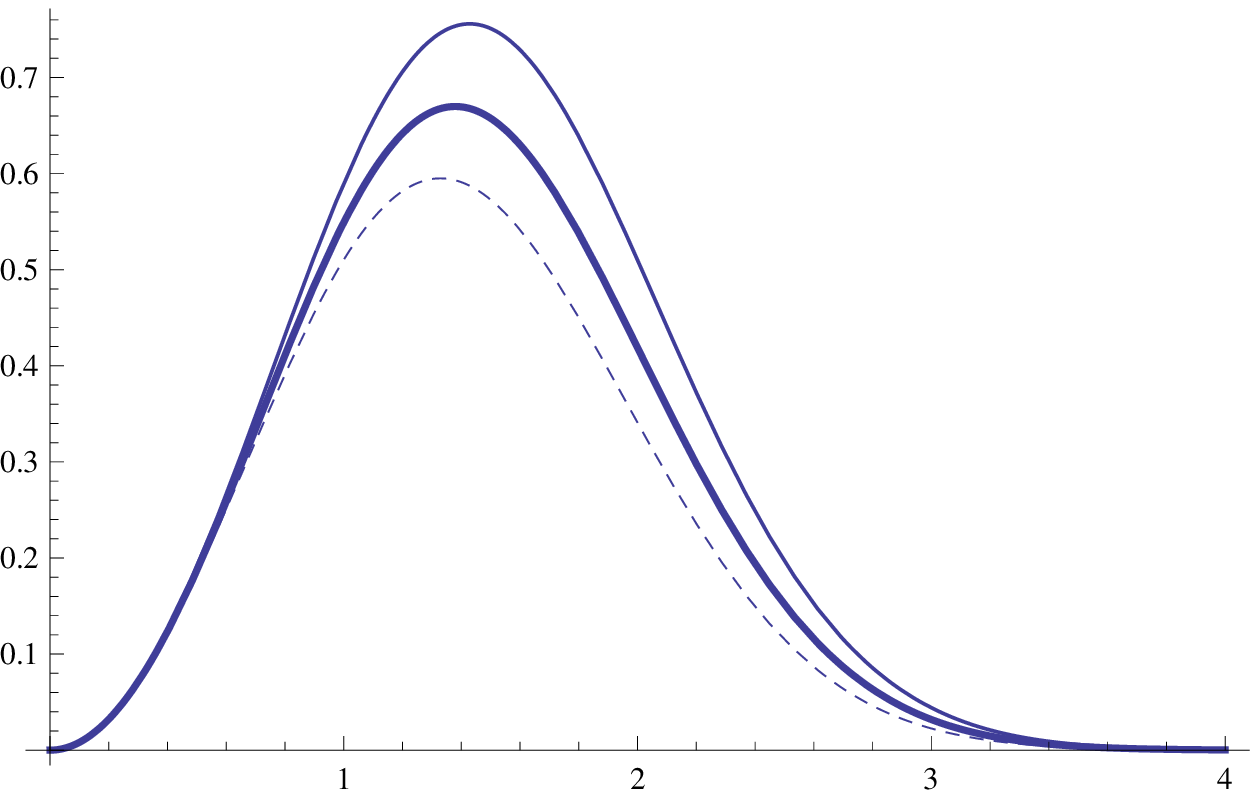}
\\
{\hspace{-2.0cm} {$y=\upsilon/\upsilon_0\longrightarrow$ }}\\
 \caption{The axion velocity distribution in the local frame. It is changing with time  due to the motion of the Earth and the sun. We exhibit  here the distribution in June  (solid line), in December  (thick solid line) and in September or March (dotted line). The last curve coincides with that in which the motion of the Earth is ignored. 
}
\label{fig:modvel}
  \end{center}
  \end{figure}

\begin{figure}[!ht]
\begin{center}
\rotatebox{90}{\hspace{-0.0cm} {$\Gamma(\alpha)/\prec\Gamma\succ\longrightarrow$}}
\includegraphics[width=0.9\textwidth,height=0.3\textwidth]{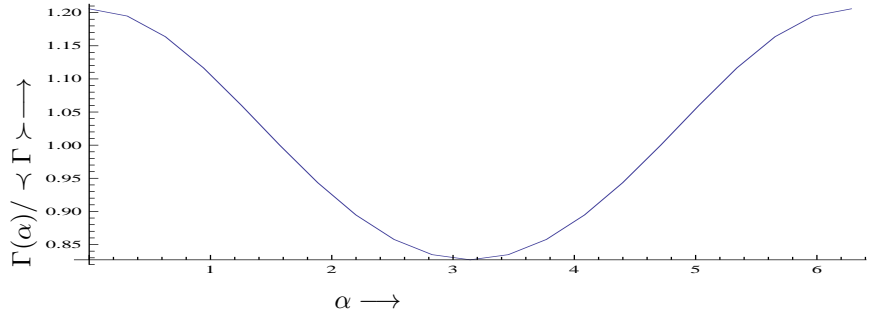}
\\
{\hspace{-2.0cm} {$\alpha\longrightarrow$ }}\\
 \caption{The ratio of the modulated width divided by that expected if the motion of the sun and the Earth is ignored, as a function of the phase of the Earth.}
\label{fig:modRW}
  \end{center}
  \end{figure}

\begin{figure}[!ht]
\begin{center}
\rotatebox{90}{\hspace{-0.0cm} {$\Gamma(\alpha)/\prec\Gamma\succ\longrightarrow$}}
\includegraphics[width=0.9\textwidth,height=0.5\textwidth]{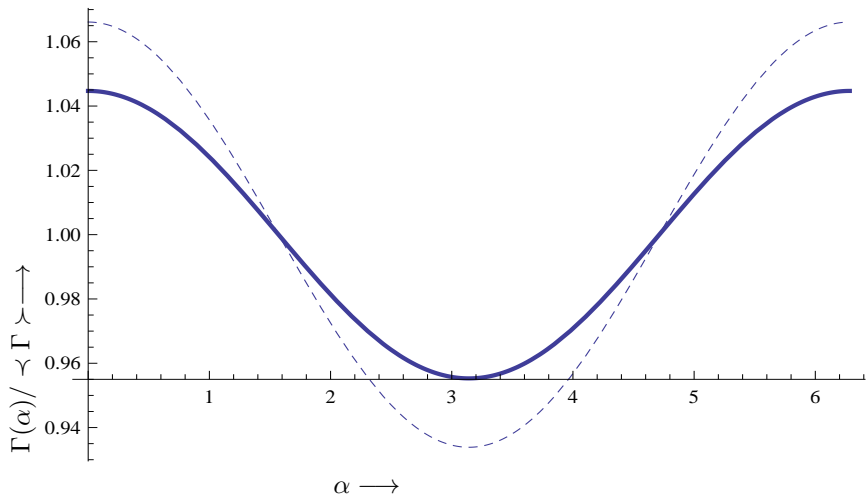}
\\
{\hspace{-2.0cm} {$\alpha\longrightarrow$ }}\\
 \caption{The ratio of the modulated width divided by the time averaged width as a function of the phase of the Earth. The solid line corresponds to the standaed M-B distribution and the dotted line to an axially symmetric M-B distribution with asymmetry parameter $\beta=0.5$ (see text).}
\label{fig:modRW2}
  \end{center}
  \end{figure}

So our next task is to transform the velocity distribution from the
galactic to the local frame. The needed equation, see e.g.
\cite{Vergados12}, is:
   \beq
{\bf y} \rightarrow {\bf y}+{\hat\upsilon}_s+\delta \left
(\sin{\alpha}{\hat x}-\cos{\alpha}\cos{\gamma}{\hat
y}+\cos{\alpha}\sin{\gamma} {\hat \upsilon}_s\right ) ,\quad
y=\frac{\upsilon}{\upsilon_0} \label{Eq:vlocal} \eeq with
$\gamma\approx \pi/6$, $ {\hat \upsilon}_s$ a unit vector in the
Sun's direction of motion, $\hat{x}$  a unit vector radially out
of the galaxy in our position and  $\hat{y}={\hat
\upsilon}_s\times \hat{x}$. The last term in the first expression
of Eq. (\ref{Eq:vlocal}) corresponds to the motion of the Earth
around the Sun with $\delta$ being the ratio of the modulus of the
Earth's velocity around the Sun divided by the Sun's velocity
around the center of the Galaxy, i.e.  $\upsilon_0\approx 220$km/s
and $\delta\approx0.135$. The above formula assumes that the
motion  of both the Sun around the Galaxy and of the Earth around
the Sun are uniformly circular. The exact orbits are, of course,
more complicated  but such deviations are not
expected to significantly modify our results. In Eq.
(\ref{Eq:vlocal}) $\alpha$ is the phase of the Earth ($\alpha=0$
around the beginning of June)\footnote{One could, of course, make the time
dependence of the rates due to the motion of the Earth more
explicit by writing $\alpha \approx(6/5)\pi\left (2 (t/T)-1 \right
)$, where $t/T$ is the fraction of the year.}.
 \\The velocity distribution in the local frame is affected by the motion of the Earth as exhibited in Fig. \ref{fig:modvel} at four characteristic periods. The ratio of the modulated width divided by that obtained by ignoring the local velocity, as a function of the phase of the Earth, is shown in Fig. \ref{fig:modRW}. 
	More appropriate in the analysis of the experiments is the relative  modulated width, i.e. the ratio of the time dependent width divided by the time averaged with, is  shown in Fig. \ref{fig:modRW2}. The results shown here are for spherically symmetric M-B distribution as well as an axially symmetric one with asymmetry parameter $\beta=0.5$ with
	 \beq
 \beta=1-\frac{\langle \upsilon_t^2\rangle}{2 \langle \upsilon_r^2\rangle}
 \eeq
with  $\upsilon_r$ the  radial, i.e. radially out of the galaxy, and $\upsilon_t$ the tangential  component of the velocity.
Essentially similar results are obtained by more exotic models, like a combination of M-B and Debris flows considered by Spegel and collaborators \cite{Spergel12}. We see that the effect is  small, around 15$\%$ difference between maximum and minimum in the presence of the asymmetry,  but still larger than that expected in ordinary dark matter searches. If we do detect the axion frequency, then we can determine its width with high accuracy and detect its modulation as a function of time.
\subsection{Asymmetry of the rates in directional experiments}
\begin{figure}[!ht]
\begin{center}
\rotatebox{90}{\hspace{-0.0cm} {$r(\Theta,\Phi)=\Gamma_dir/\Gamma_{st}\rightarrow$}}
\includegraphics[width=0.9\textwidth,height=0.5\textwidth]{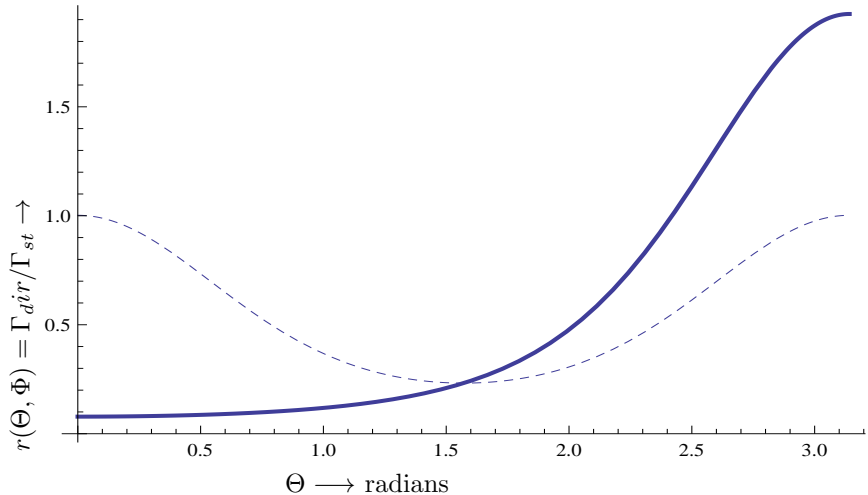}\\
{\hspace{-2.0cm} {$\Theta\longrightarrow$ radians}}\\
 \caption{The  ratio of the width expected in a directional experiment divided by that expected in a standard experiment. The solid line is expected, if the sense of direction is known, while the dotted will show up, if the sense of direction is not known.}
 \label{fig:Axionwidth}
  \end{center}
  \end{figure}

Consideration of the velocity distribution will give an important signature, if directional experiments become feasible. This can be seen as follows:
\begin{itemize}
\item The width will depend  specified by two angles $\Theta$ and 
$\Phi$. \\
The angle $\Theta$  is the polar angle  between the sun's velocity and the direction of observation. The angle $\Phi$ is measured in a plane perpendicular to the sun's velocity, starting from the line coming radially out of the galaxy and passing through the sun's location.
\item The axion velocity, in units of the solar velocity, is given as 
\beq
{\bf y}=y\left ({\hat x}\sqrt{1-\xi^2}\cos{\phi}+ {\hat y}\sqrt{1-\xi^2}\sin{\phi}+{\hat z}\xi \right  )
\eeq
\item Set $\delta=0$, i.e. ignore the motion of the Earth around the sun.\\
Then the velocity distribution in the local frame is obtained by the substitution:
\barr
\upsilon^2 \rightarrow&&\upsilon_0^2\left(y^2 +1+2y\left (\xi \cos{\Theta}+\sqrt{1-\xi^2}\sin{\Theta}\left (\cos{\Phi}\cos{\phi} \right .  \right . \right .\nonumber\\
 &+& \left . \left . \left .\sin{\Phi}\sin{\phi}\right ) \right )\right )
\earr
One then can integrate over $\xi$ and $\phi$. The results become essentially independent of $\Phi$, so long as the motion of the Earth around the sun is ignored\footnote{The annual modulation of the expected results  due to the motion of the Earth around the sun will show up  in  the directional experiments as well, but it is going to be less important and it will not be discussed here.}.Thus we  obtain $\prec \upsilon^2\succ$ from the  axion velocity distribution for various polar angles $\Theta$. 
\end{itemize}
We write the width observed in a directional experiment as:
\beq
\Gamma=r(\Theta)\Gamma_{st}
\eeq
where $\Gamma_{st}$ is the width in the standard experiments. Ignoring the motion of the Earth around the sun the factor $r$ depends only on $\Theta$. Furthermore, if for simplicity we ignore the upper velocity bound (cut off)  in the M-B distribution, i.e. the escape velocity $\upsilon_{esc}=2.84 \upsilon_0$, we can get the solution in analytic form. We find:
\beq
r(\Theta)=\frac{2}{5}\frac {e^{-1}}{2}\left ( e^{\cos ^2{\Theta }} (\cos {2 \Theta }+4) \mbox{ erfc}(\cos \Theta
   )-\frac{2 \cos \Theta }{\sqrt{\pi }}\right ),\, \mbox{(sense  known)}
\eeq
where 
$$\mbox{erfc}(z)=1-\mbox{erf}(z), \mbox{erf}(z)=\int_0^z dt e^{-t^2} \mbox{( error function)}.$$
\beq
r(\Theta)=\frac{2}{5}\frac{1}{2} e^{-\sin ^2\Theta } (\cos{ 2 \Theta }+4),\, \mbox{(sense of direction not known)}
\eeq 
The adoption of an upper cut off has little effect. In Fig. \ref{fig:Axionwidth} we present the exact results.
	 The above results were obtained with a M-B velocity distribution\footnote{Evaluation  of the relevant average velocity squared  in  some other  models \cite{Sikivie11},\cite{GHP-W14}, which lead to caustic ring distributions, can also be worked out for axions as above in a fashion
 analogous to that of WIMPs  \cite{Verg01}, but this is not the subject of the present paper}. 

Our  results  indicate that the width will exhibit diurnal variation! For a cylinder of Length $L$ such a variation is expected to be favored \cite{IrasGarcia12} in the regime of $m_a L=10-25\times 10^{-4}$eV-m.
	This diurnal variation will be discussed in the next section.
	
	\section{The diurnal variation in directional experiments}
The apparatus will be oriented in a direction specified in the local frame, e.g. by a point in the sky specified, in the equatorial system, by right ascension $\tilde{\alpha}$ and inclination $\tilde{\delta}$\footnote{We have chosen to adopt the notation $\tilde{\alpha}$ and $\tilde{\delta}$ instead of the standard notation $\alpha$ and $\delta$ employed by the astronomers to avoid possible confusion stemming from the fact that $\alpha$  is usually  used to designate the phase of the Earth and $\delta$  for the ratio of the rotational velocity of the Earth around the Sun  by the velocity of the sun around the center of the galaxy}. This will lead to a diurnal variation\footnote{This should not be confused with the diurnal variation expected even in non directional experiments due to the rotational velocity of the Earth, which is expected to be too small.} of the event rate \cite{CYGNUS09}. This situation has already been discussed in the case of standard WIMPs \cite{VerMou11}. We will briefly discuss the transformation into the relevant astronomical coordinates here.

The galactic frame, in the so called J2000 system, is defined by the galactic pole with ascension $\tilde{\alpha}_1=12^h~ 51^m~ 26.282^s$ and inclination $\tilde{\delta}_1= +27^0 ~7^{'} ~42.01^{''}$ and the galactic center at $\tilde{\alpha}_2 =17^h ~45^m~ 37.224^s$ , $\tilde{\delta}_2=-(28^0~ 56^{'}~ 10.23^{''} ) $. Thus the galactic unit vector $\hat{y}$, specified by $(\tilde{\alpha}_1,\tilde{\delta}_1)$, and the unit vector $\hat{s}$, specified by $(\tilde{\alpha}_2,\tilde{\delta}_2)$,
can be expressed in terms of the celestial unit vectors $\hat{i}$ (beginning of measuring the right ascension),  $\hat{k}$ (the axis of the Earth's rotation) and $\hat{j}=\hat{k}\times\hat{i}$.
One finds
 \barr
{\hat y}&=&-0.868{\hat i}- 0.198{\hat j}+0.456{\hat k}
\nonumber \mbox{ (galactic axis) },\\
{\hat x}&=&-{\hat s}=0.055{\hat i}+ 0.873{\hat j}+ 0.483{\hat k} \mbox{ (radially out to the sun)  },
\nonumber\\
{\hat z}&=&{\hat x}\times{\hat y}=0.494{\hat i}- 0.445{\hat j} +0.747{\hat k}\mbox{ ( sun's velocity)}.
\earr
Note in our system the x-axis is opposite to the s-axis used by the astronomers.
Thus a vector oriented by $(\tilde{\alpha},\tilde{\delta}) $ in the laboratory  is given   in the galactic frame by a unit vector with components:
\beq
\left (
\begin{array}{l}
 y \\
 x \\
 z
\end{array}
\right )
=\left (\begin{array}{l}
 -0.868 \cos {\tilde{\alpha} } \cos {\tilde{\delta} }-0.198 \sin {\tilde{\alpha}
   } \cos {\tilde{\delta }}+0.456 \sin{\tilde{\delta} } \\
 0.055 \cos {\tilde{\alpha}} \cos {\tilde{\delta}}+0.873 \sin {\tilde{\alpha}
   } \cos {\tilde{\delta }}+0.4831 \sin {\tilde{\delta }} \\
 0.494 \cos {\tilde{\alpha} } \cos {\tilde{\delta} }-0.445 \sin {\tilde{\alpha}
   } \cos {\tilde{\delta}}+0.747 \sin {\tilde{\delta} }
\end{array}
\right ).
\eeq
This can also be parametrized  as:
\barr
x&=&\cos {\gamma } \cos {\tilde{\delta} } \cos \left(\tilde{\alpha} -\tilde{\alpha} _0\right)-\sin
   {\gamma } \left(\frac{}{}\cos {\tilde{\delta} } \cos{\theta _P} \sin
   \left(\tilde{\alpha} -\tilde{\alpha} _0\right)\right .\nonumber\\
	&+& \left . \sin {\tilde{\delta} } \sin{\theta
   _P}\right),
\earr
\beq
y=\cos \left(\theta _P\right) \sin {\tilde{\delta} }-\cos {\tilde{\delta} } \sin
   \left(\tilde{\alpha} -\tilde{\alpha} _0\right) \sin{\theta _P},
\eeq
\barr
z&=&\cos {\delta } \cos \left(\tilde{\alpha} -\tilde{\alpha} _0\right) \sin {\gamma }+\cos
   {\gamma } \left(\frac{}{}\cos {\tilde{\delta} } \cos {\theta _P} \sin
   \left(\tilde{\alpha} -\tilde{\alpha} _0\right)\right . \nonumber\\
	&+& \left . \sin {\tilde{\delta} } \sin{\theta
   _P}\right),
\earr
where $\tilde{\alpha}_0=282.25^0$ is the right ascension of the equinox, $\gamma\approx 33^0$ was given above and $\theta_P=62.6^0$ is the angle the Earth's north pole forms with the axis of the galaxy. \\Due to the Earth's rotation the unit vector $(x,y,z)$, with a suitable choice of the initial time, $\tilde{\alpha}-\tilde{\alpha}_0=2\pi(t/T)$, is changing as a function of time

\barr
x&=&\cos {\gamma } \cos {\tilde{\delta} } \cos \left(\frac{2 \pi  t}{T}\right)-\sin
   {\gamma } \left(\frac{}{}\cos {\delta } \cos{\theta _P} \sin
   \left(\frac{2 \pi  t}{T}\right)\right .\nonumber\\
	&+& \left . \sin {\tilde{\delta} } \sin{\theta
   _P}\right),
\earr

\beq
y=\cos \left(\theta _P\right) \sin {\tilde{\delta }}-\cos {\tilde{\delta} } \sin
   \left(\frac{2 \pi  t}{T}\right) \sin{\theta _P},
\eeq

\barr
z&=&\cos \left(\frac{2 \pi  t}{T}\right) \cos {\tilde{\delta} } \sin {\gamma }\nonumber\\
&+&\cos
   {\gamma } \left(\cos {\tilde{\delta} } \cos{\theta _P} \sin
   \left(\frac{2 \pi  t}{T}\right)
	+ \sin {\tilde{\delta} } \sin{\theta
   _P}\right ),
\earr
where $T$ is the period of the Earth's rotation.

Some  points of interest are:
\barr
&&\mbox{The celestial pole: }\nonumber\\
 (y,x,z)&=&(0.460,0.484,0.745)\Rightarrow(\theta=62.6^0,\phi=57^0),
\nonumber\\
&&\mbox{The ecliptic pole: }\nonumber\\
 (y,x,z)&=&(0.497,0.096,0.863)\Rightarrow(\theta=62.6^0,\phi=83.7^0),
\nonumber\\
&&\mbox{The equinox: }\nonumber\\
 (y,x,z)&=&(-0.868,0.055,0.494)\Rightarrow(\theta=150.2^0,\phi=83.7^0).\nonumber\\
\earr
where $\theta$ is defined with respect to the polar axis (here $y$) and $\phi $ is measured from the $x$ axis towards the $z$ axes.

 Thus the angles $\Theta$, which is of interest to us  in directional experiments, is given by
\beq
\Theta=\cos^{-1}{z},
\eeq
An analogous, albeit a bit more complicated expression can be derived for the angle  $\Phi$.

\begin{figure}[!ht]
 \begin{center}
\rotatebox{90}{\hspace{-0.0cm} {$\Theta  \longrightarrow$ radians}}
     \includegraphics[width=0.9\textwidth,height=0.5\textwidth]{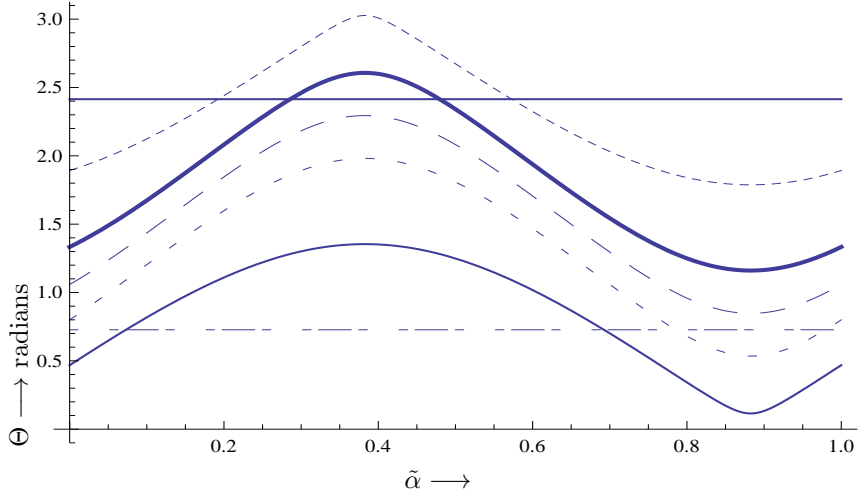}\\
\hspace{-0.0cm} {$\tilde{\alpha}\longrightarrow$}
\caption{ Due to the diurnal motion of the Earth different angles $\Theta$ in galactic coordinates are sampled as the earth rotates. The angle $\Theta$  scanned by the direction of observation is shown for various inclinations $\tilde{\delta}$. 
We see that, for negative inclinations, the angle $\Theta$ can take values near $\pi$, i.e. opposite to the direction of the sun's velocity, where the rate attains its maximum.For an explanation of the curves see Fig. \ref{fig:diurnal}}
 \label{fig:theta}
  \end{center}
  \end{figure}

 The  angle $\Theta$  scanned by the direction of observation is shown, for various inclinations $\tilde{\delta}$, in Fig.~\ref{fig:theta}. We see that for  negative inclinations, the angle $\Theta$ can take values near $\pi$, i.e. opposite to the direction of the sun's velocity, where the rate attains its maximum (see Fig. \ref{fig:theta}).   

\begin{figure}[!ht]
 \begin{center}
\subfloat[]
{
{\rotatebox{90}{\hspace{-0.0cm} {$\Gamma/\Gamma_{st}$} (sense known)}}
\includegraphics[width=0.9\textwidth,height=0.5\textwidth]{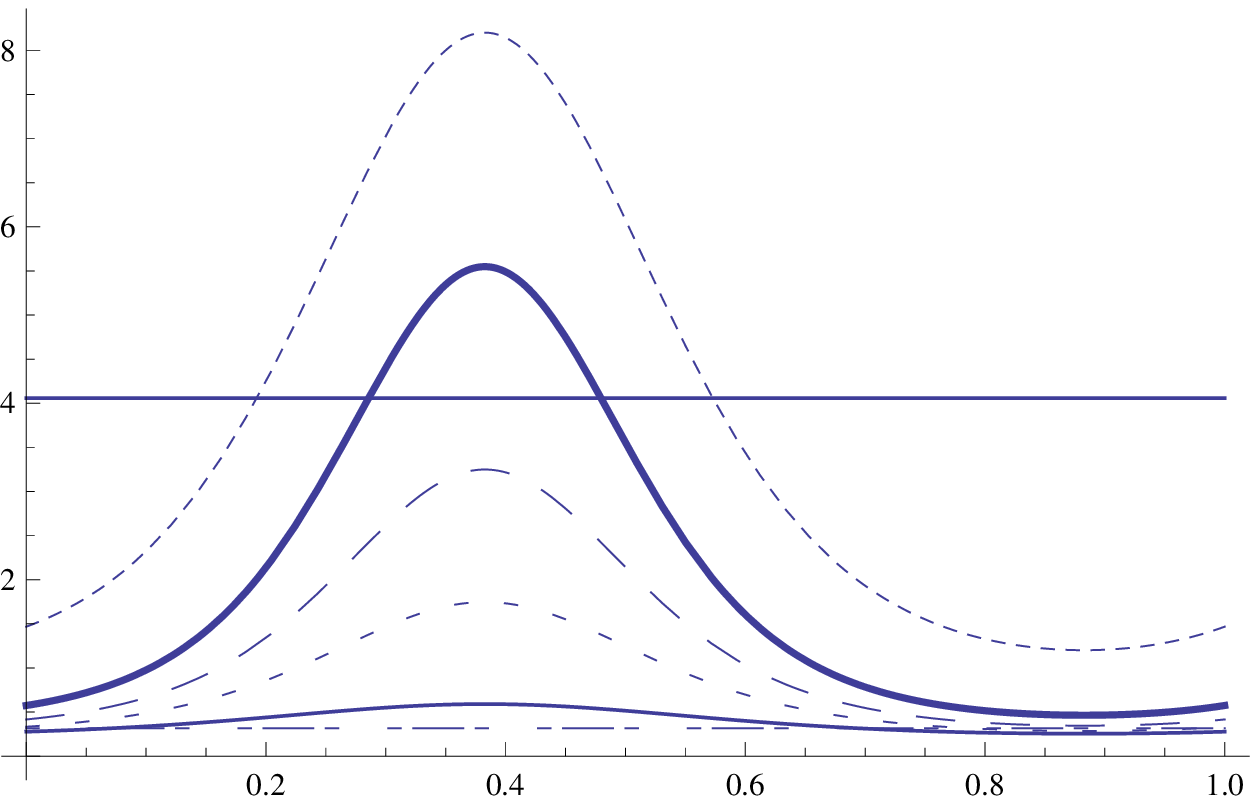}
\hspace{-0.0cm} {$\frac{t}{T} \longrightarrow$}
}\\
{\subfloat[]
{\rotatebox{90}{\hspace{-0.0cm} {$\Gamma/\Gamma_{st}$} (sense unknown)}}
\includegraphics[width=0.9\textwidth,height=0.5\textwidth]{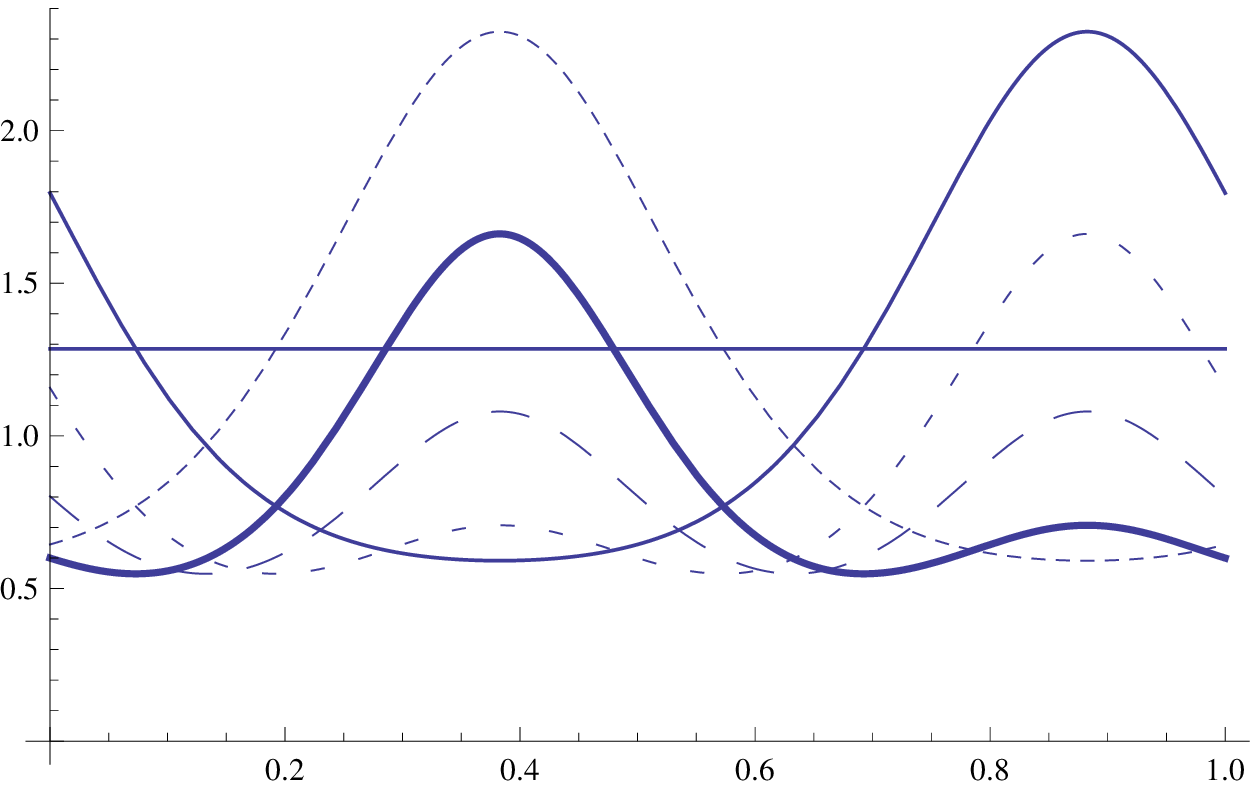}
\hspace{-0.0cm} {$\frac{t}{T} \longrightarrow$}
}
\\
\caption{The time dependence (in units of the Earth's rotation period)   of the ratio of the directional width divided by the non directional width for various inclinations $\tilde{\delta}$, when the sense can be determined (a) or both senses are included (b). In The curves indicated by intermediate thickness solid, the short dash, thick solid line, long dashed, dashed, fine solid line,  and the long-short dashed correspond to inclination $\tilde{\delta}=-\pi/2,-3\pi/10,-\pi/10,0,\pi/10,3\pi/10$ and $\pi/2$  respectively. We see that, for negative inclinations, the angle $\Theta$ can take values near $\pi$, i.e. opposite to the direction of the sun's velocity, where the rate attains its maximum if the sense of direction is known. There is no time variation, of course, when $\tilde{\delta}=\pm\pi/2$. }
 \label{fig:diurnal}
  \end{center}
  \end{figure}

The equipment scans different parts of the galactic sky, i.e. observes different angles $\Theta$. So the rate will change with time depending on whether the sense of of observation. We assume that the sense of direction can be distinguished in the experiment. 
The total flux  is exhibited in Fig. \ref{fig:diurnal}.
	\section{Discussion}
	In the present work we discussed the time variation of the width of of the axion to photon resonance cavities involved in Axion Dark Matter Searches. We find two important signatures:
	\begin{itemize}
	\item Annual variation due to the motion of the Earth around the sun. We find that  in the relative width, i.e. the width divided by its time average, can attain differences of  about $15\%$ between the maximum expected in June  and the minimum  expected six months later. This variation is larger than the modulation expected  in ordinary dark matter of WIMPs. It does not depend on the geometry of the cavity or other details of the apparatus. It does not depend strongly on the assumed velocity distribution.
	\item A characteristic diurnal variation in of the width in directional experiments with most favorable scenario in the range of $m_e L=1.0-2.5\times 10^{-3}$eV\,m. This arises from asymmetries of the local axion velocity with respect to the sun's direction of motion manifested in a time dependent way due to the rotation of the Earth around its own axis. Admittedly such experiments are much  harder, but the expected signature persists, even if one cannot tell the direction of motion of the axion velocity entering in the expression of the width.  Anyway once such a device is operating, data  can be taken  as usual. Only one has to  bin them according the  time they were obtained. If a potentially useful signal is found,  a complete analysis can be done according the directionality to firmly establish that the signal is due to the axion.
	\end{itemize}
	In conclusion in this work we have elaborated on two signatures that might aid the analysis of axion dark matter searches.
	
{\bf	Acknowledgments}: One of the authors (JDV) is indebted to Professor J. E. Kim for useful discussions and to Leslie Rosenberg for his   careful reading of the manuscript and his useful comments on directional experiments. IBS-Korea partially supported this project under system code IBS-R017-D1-2014-a00.

\begin{thebibliography}{10}
\expandafter\ifx\csname url\endcsname\relax
  \def\url#1{\texttt{#1}}\fi
\expandafter\ifx\csname urlprefix\endcsname\relax\def\urlprefix{URL }\fi
\expandafter\ifx\csname href\endcsname\relax
  \def\href#1#2{#2} \def\path#1{#1}\fi

\bibitem{PecQui77}
R.~Peccei, H.~Quinn, Phys. Rev. Lett 38 (1977) 1440.

\bibitem{SWeinberg78}
S.~Weinberg, Phys. Rev. Lett. 40 (1978) 223.

\bibitem{Wilczek78}
F.~Wilczek, Phys. Rev. Lett. 40 (1978) 279.

\bibitem{PWW83}
J.~Preskill, M.~B. Wise, F.~Wilczek, Phys. Lett. B120 (1983) 127.

\bibitem{AbSik83}
L.~F. Abbott, P.~Sikivie, Phys. Lett. B120 (1983) 133.

\bibitem{DineFisc83}
M.~Dine, W.~Fischler, Phys. Lett. B120 (1983) 137.

\bibitem{PriSecSad88}
J.~Primack, D.~Seckel, B.~Sadoulet, Ann. Rev. Nuc. Par. Sc. 38 (1988) 751.

\bibitem{Sikivie83}
P.~Sikivie, Phys. Rev. Lett. 51 (1983) 1415.

\bibitem{Stern14}
I. P. Stern, ArXiv 1403.5332 (2014) physics.ins--det, on behalf of ADMX and
  ADMX-HF collaborations, Axion Dark Matter Searches.

\bibitem{MultIBSExp}
G. Rybka, The Axion Dark Matter Experiment, IBS MultiDark Joint Focus Program
  WIMPs and Axions, Daejeon, S. Korea October 2014.

\bibitem{ExpSetUp11b}
S.~J. Asztalos, et~al., Phys. Rev. Lett. 104 (2010) 041301, the ADMX
  Collaboration, arXiv:0910.5914 (astro-ph.CO).

\bibitem{ADMX10}
A.~Wagner, et~al., Phys. Rev. Lett. 105 (2010) 171801, for the ADMX
  collaboration; arXiv:1007.3766 (astro-ph.CO).

\bibitem{CAPP}
Center for Axion and Precision Physics research (CAPP), Daejeon 305-701,
  Republic of Korea. More information is available at http:$/
  /\mbox{capp.ibs.re.kr}/\mbox{html}/\mbox{capp}\_\mbox{en}/$.

\bibitem{ExpSetUp11a}
S.~J. Asztalos, et~al., Nucl. Instr. Meth. in Phys. Res. A656 (2011) 39,
  arXiv:1105.4203 (physics.ins-det).

\bibitem{MultIBSTh}
G. Raffelt, Astrophysical Axion Bounds , IBS MultiDark Joint Focus Program
  WIMPs and Axions, Daejeon, S. Korea October 2014.

\bibitem{HKNS14}
J. Hong, J. E. Kim, S. Nam, Y. Semertzidis, arXiv:1403.1576 (2014)
  physics.ins--det, calculations of Resonance enhancement factor in
  axion-search tube experiments.

\bibitem{JEKim98}
J.~E. Kim, Phys. Rev. D 58 (1998) 055006.

\bibitem{KMWM85}
L.~Krauss, J.~Moody, F.~Wilczek, D.~Morris, Phys. Rev. Lett. 55 (1985) 1797.

\bibitem{FutExp14}
T. M. Shikair {\it et al}, Future Directions in Microwave Search for Dark
  Matter Axions, arXiv:1405.3685 [physics.ins-det], (to appear in IJMPA).

\bibitem{IrasGarcia12}
I.~G. Irastorza, J.~A. García, JCAP 1210 (2012) 022, arXiv:1007.3766
  (astro-ph.IM).

\bibitem{DFS86}
A.~K. Drukier, K.~Freese, D.~N. Spergel, Phys. Rev. D 33 (1986) 3495.

\bibitem{Turner90}
M.~Turner, Phys. Rev. D 42 (1990) 3572.

\bibitem{Vergados12}
J.~Vergados, Phys. Rev. D. 85 (2012) 123502, ; arXiv:1202.3105 (hep-ph).

\bibitem{Spergel12}
M.~Kuhlen, M.~Lisanti, D.~Speregel, Phys. Rev. D 86 (2002) 063505,
  arXiv:1202.0007 (astro-ph.GA).

\bibitem{Sikivie11}
P.~Sikivie, Phys. Lett. B695 (2011) 22.

\bibitem{GHP-W14}
A. H. Guth , M. P. Hertzberg and C. Prescond-Weistein,arXiv:1412.5930
  [astro-ph.CO].

\bibitem{Verg01}
J.~D. Vergados, Phys. Rev. D 63 (2001) 063511.

\bibitem{CYGNUS09}
S. Ahlen, The case for a directional dark matter detector and the status of
  current experimental efforts, cygnus2009Whitepaper, Edited by J. B. R.
  Battat, IJMPA 25 (20010) 1; arXiv:0911.0323 (astro-ph.CO).

\bibitem{VerMou11}
J.~D. Vergados, C.~C. Moustakidis, Eur. J. Phys. 9(3) (2011) 628,
  arXiv:0912.3121 [astro-ph.CO].

\end{thebibliography}

\end{document}